\documentclass[manuscript]{emulateapj}

\shorttitle{Blowout jet in AR 12192} \shortauthors{Li et al.}

\journalinfo{Accepted for publication in ApJL}

\submitted{Accepted for publication in ApJL}

\begin{document}

\title{Trigger of a blowout jet in a solar coronal mass ejection associated with a flare}

\author{Xiaohong Li, Shuhong Yang, Huadong Chen, Ting Li, Jun Zhang}

\affil{Key Laboratory of Solar Activity, National
Astronomical Observatories, Chinese Academy of Sciences, Beijing
100012, China; shuhongyang@nao.cas.cn}

\begin{abstract}

Using the multi-wavelength images and the photospheric magnetograms
from the \emph{Solar Dynamics Observatory}, we study the flare which
was associated by the only one coronal mass ejection (CME) in active
region (AR) 12192. The eruption of a filament caused a blowout jet,
and then an M4.0 class flare occurred. This flare was located at the
edge of AR instead of in the core region. The flare was close to the
apparently ``open" fields, appearing as extreme-ultraviolet
structures that fan out rapidly. Due to the interaction
between flare materials and ``open" fields, the flare
became an eruptive flare, leading to the CME. Then at the same site
of the first eruption, another small filament erupted. With the high
spatial and temporal resolution H$\alpha$ data from the New Vacuum
Solar Telescope at the \emph{Fuxian Solar Observatory}, we
investigate the interaction between the second filament and the
nearby ``open" lines. The filament reconnected with the ``open"
lines, forming a new system. To our knowledge, the detailed process
of this kind of interaction is reported for the first time. Then the
new system rotated due to the untwisting motion of the filament,
implying that the twist was transferred from the closed filament
system to the ``open" system. In addition, the twist seemed to
propagate from the lower atmosphere to the upper layers, and was
eventually spread by the CME to the interplanetary space.

\end{abstract}


\section{Introduction}

Solar jets which are small-scale plasma ejections along open field
lines or the legs of large-scale coronal loops have been extensively
investigated (e.g., Schmieder et al. 1988; Shibata et al. 1994;
Zhang et al. 2000; Liu \& Kurokawa 2004; Cirtain et al. 2007; Jiang
et al. 2007; Tian et al. 2014). They often recur from the regions of
evolving photospheric magnetic flux, such as satellite spots of an
active region (AR), quiet regions, and coronal holes (e.g., Chae et
al. 1999; Chen et al. 2008; Pariat et al. 2010; Yang et al. 2011a,
2011b). The detailed statistical properties of X-ray jets were
studied by Shimojo et al. (1996) and Savcheva et al. (2007).
Probably due to the release of magnetic twist, helical or untwisting
motions are often observed in jets (e.g., Jibben \& Canfield 2004;
Liu et al. 2009; Shen et al. 2011; Chen et al. 2012). Sometimes,
jets are associated with large-scale solar eruptions, such as
filament eruption and coronal mass ejections (CMEs; e.g. Liu et al.
2005; Jiang et al. 2008; Guo et al. 2010;  Shen et al. 2012).

By examining many X-ray jets in \emph{Hinode}/X-Ray Telescope
coronal X-ray movies of the polar coronal holes, Moore et al. (2010)
found that nearly all solar polar X-ray jets can be divided into two
different types: standard jets and blowout jets. The standard X-ray
jets are evidently produced by reconnection between an emerging
magnetic arch at the base of the jet and the ambient unipolar open
magnetic field of the coronal hole, during which the interior of the
base arch is insert and does not participate in the eruption (e.g.,
Yokoyama \& Shibata 1995; Canfield et al. 1996; Pariat et al. 2009).
The blowout X-ray jets are produced by blowout eruption of the base
small filament, a miniature version of the blowout eruption of the
sheared-core magnetic arcade in a CME eruption. The blowout jet
model has been supported by some observational studies (e.g., Hong
et al. 2011; Liu et al. 2011; Shen et al. 2012; Pucci et al. 2013;
Young \& Muglach 2014). However, the detailed process of the
interaction between the erupting filament and the ambient open field
is unclear.

The super solar AR 12192 during 2014 October consisted of the
largest sunspot group in the past 24 years (e.g., Sun et al. 2015;
Thalmann et al. 2015; Yang et al. 2015). It is so far the most
intensely flaring region of Cycle 24. According to the statistics of
Chen et al. (2015), AR 12192 produced 6 X-class and 29 M-class
flares from October 18 to 29, but only one of which was followed by
a CME. In this study, using the observations of the New Vacuum Solar
Telescope (NVST; Liu et al. 2014) and the \emph{Solar Dynamics
Observatory} (\emph{SDO}; Pesnell et al. 2012), we investigated the
only one CME-associated flare from the periphery of AR 12192, which
was caused by a blowout jet (Chen et al. 2015). Our observations
present the unprecedentedly clear process of the interaction between
the erupting small filament and the ambient ``open" field, which is
consistent with the blowout jet model. Here and in the
following text, we use quotation marks around the word ``open",
considering that the ``open" loops, which appear as rapidly fan-out
extreme-ultraviolet (EUV) structures, might be large-scale loops
reaching large heights as calculated by Thalmann et al. (2015).

\section{Observations and data analysis}

\begin{figure*}
\centering
\includegraphics
[bb=39 255 531 571,clip,angle=0,width=0.8\textwidth]{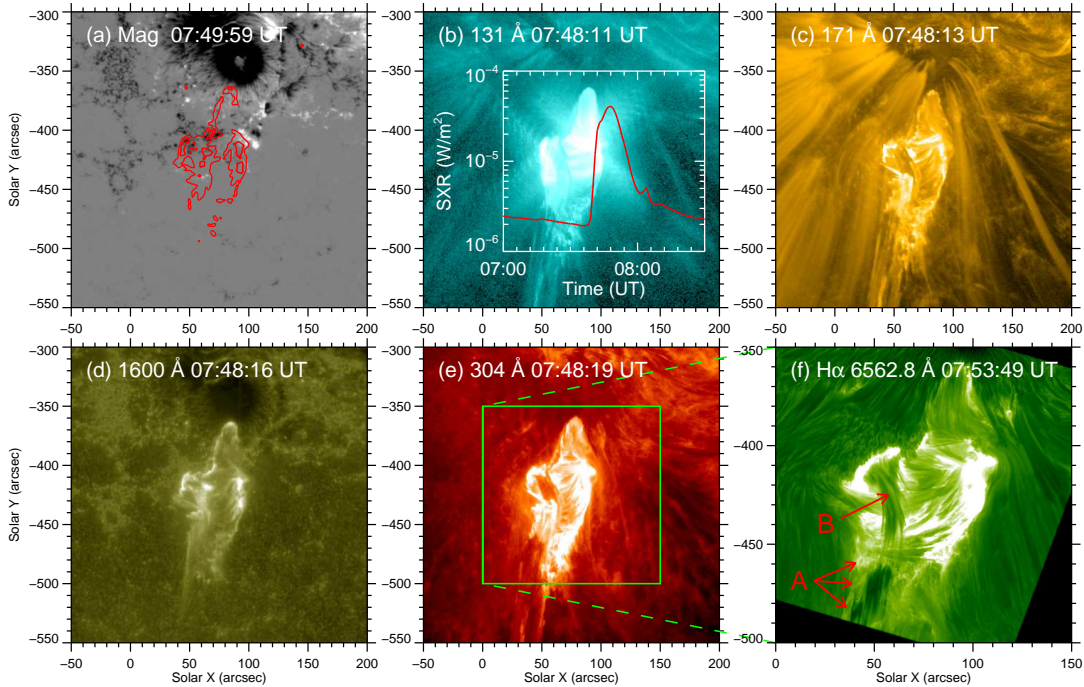}
\caption{HMI LOS magnetogram (panel (a)), AIA multi-wavelength
images (panels (b)-(e)), and NVST H$\alpha$ image (panel (f))
displaying the appearance of the first eruption. The red
contours in panel (a) display the flare ribbons in 1600 {\AA}.
In panel (b), the red curve displays the variation of the
GOES soft X-ray flux. The green square in panel (e) outlines the
FOV of panel (f). Arrows ``A" and ``B" in panel (f) denote
the jet and the filament, respectively. \label{fig1}}
\end{figure*}

\begin{figure*}
\centering
\includegraphics
[bb=108 242 459 582,clip,angle=0,width=0.8\textwidth]{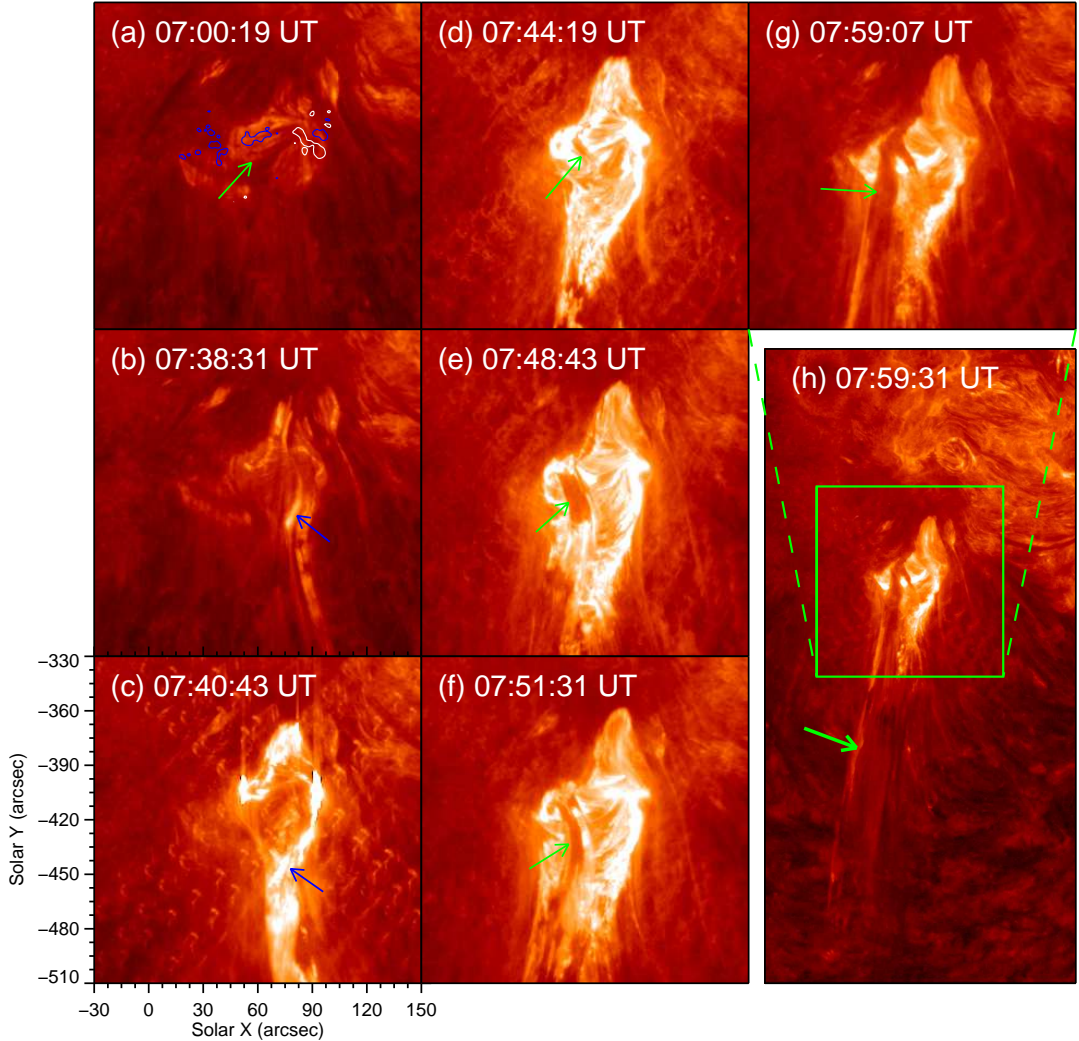}
\caption{ AIA 304 {\AA} images showing the first eruption (also see
the animation movie\_1.mp4). The white and blue curves in panel (a)
are the contours of the positive and negative polarity magnetic
fields, respectively. In panels (a)-(g), the green arrows denote the
filament and the blue arrows indicate the blowout jet. In panel (h),
the arrow denotes the filament material ejected outward.
\label{fig2}}
\end{figure*}

\begin{figure*}
\centering
\includegraphics
[bb=85 263 464 561,clip,angle=0,width=0.8\textwidth]{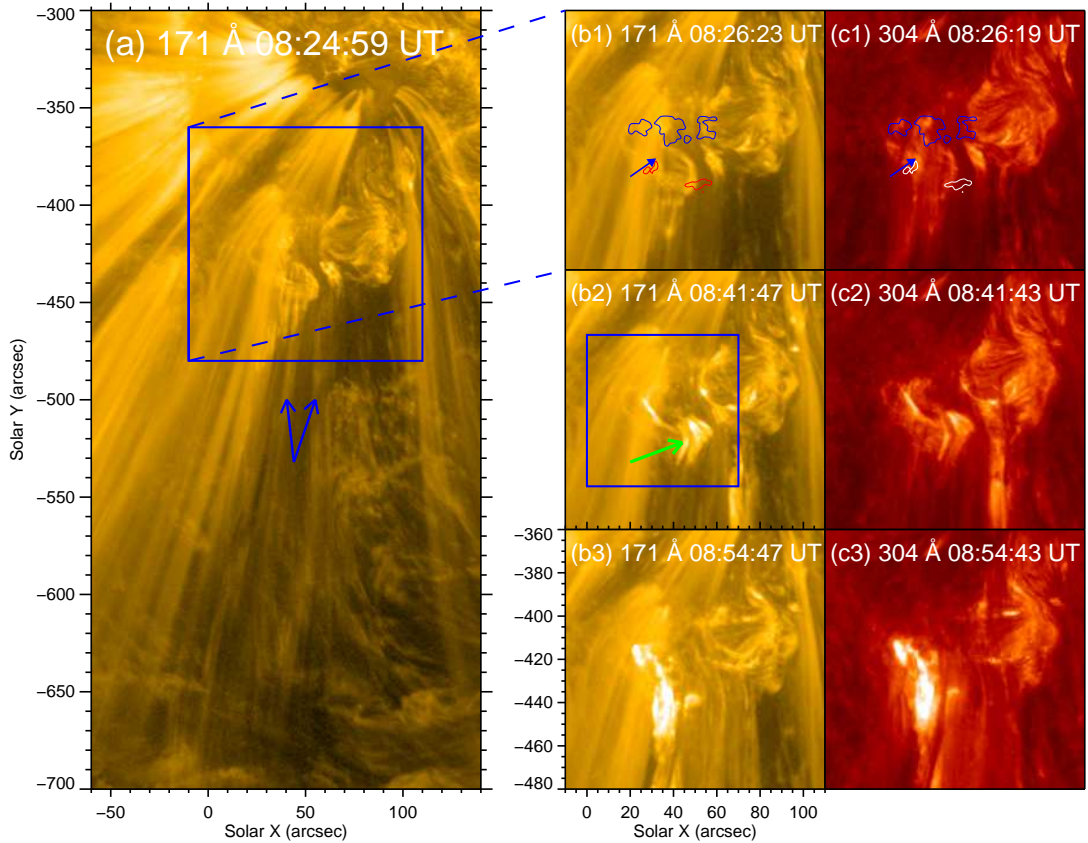}
\caption{AIA 171 {\AA} image (panel (a)) displaying the general
appearance and the surrounding environment before the second
eruption. AIA 171 {\AA} (panels (b1)-(b3)), and 304 {\AA} (panels
(c1)-(c3)) images showing the second eruption process (also
see the animation movie\_2.mp4). The white and red curves are the
contours of the positive polarity fields, and the blue curves are
the contours of the negative fields. The arrows in panel (a) denote
the ``open" field lines, and the arrows in panels (b1) and (c1)
indicate the filament. The green arrow in panel (b2)
indicates the brightening region. The box in panel (a) outlines the
FOV of panels (b1)-(c3), and the box in panel (b2) mark the FOV of
Figure 4. \label{fig3}}
\end{figure*}

\begin{figure*}
\centering
\includegraphics
[bb=62 222 540 605,clip,angle=0,width=0.8\textwidth]{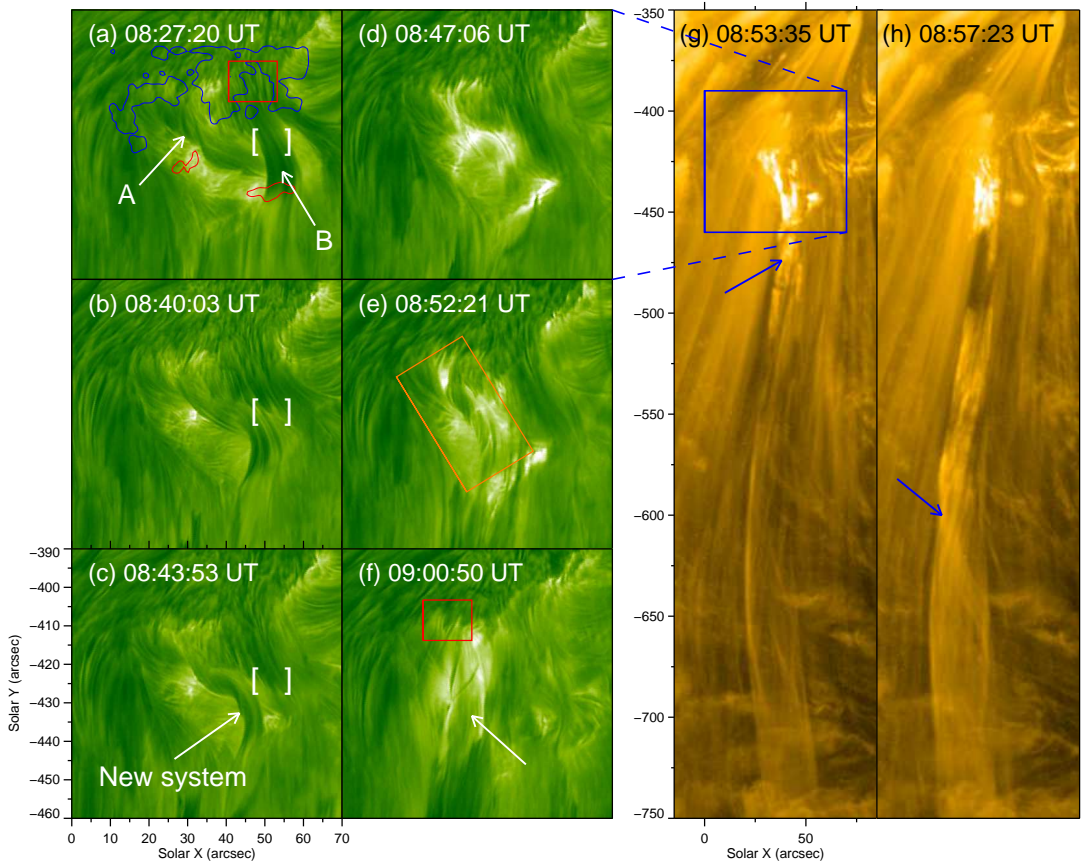}
\caption{Panels (a)-(f): NVST H$\alpha$ images showing the detailed
interaction process of the filament and the ``open" lines (also see
the animation movie\_3.mp4). Panels (g)-(h): AIA 171 {\AA} images
displaying the outward propagation process of the magnetic
twist (also see the animation movie\_4.mp4). The red and
blue curves in panel (a) are the contours of the positive and
negative magnetic fields, respectively. Arrows ``A" and ``B" in
panel (a) denote the filament and the ``open" lines, respectively.
The arrow in panel (c) indicates the new system, and the arrow in
panel (f) denotes the structure of the system after rotation. The
brackets in panels (a)-(c) mark the area where the ``open" field
lines disconnected. The red rectangles in panels (a) and (f) outline
the footpoints of the ``open" lines. The brown rectangle in panel
(e) outlines the FOV of Figure 5. The arrows in panels (g) and (h)
indicate the positions of the outward propagating twist.
\label{fig4}}
\end{figure*}

\begin{figure*}
\centering
\includegraphics
[bb=82 174 502 661,clip,angle=0,width=0.8\textwidth]{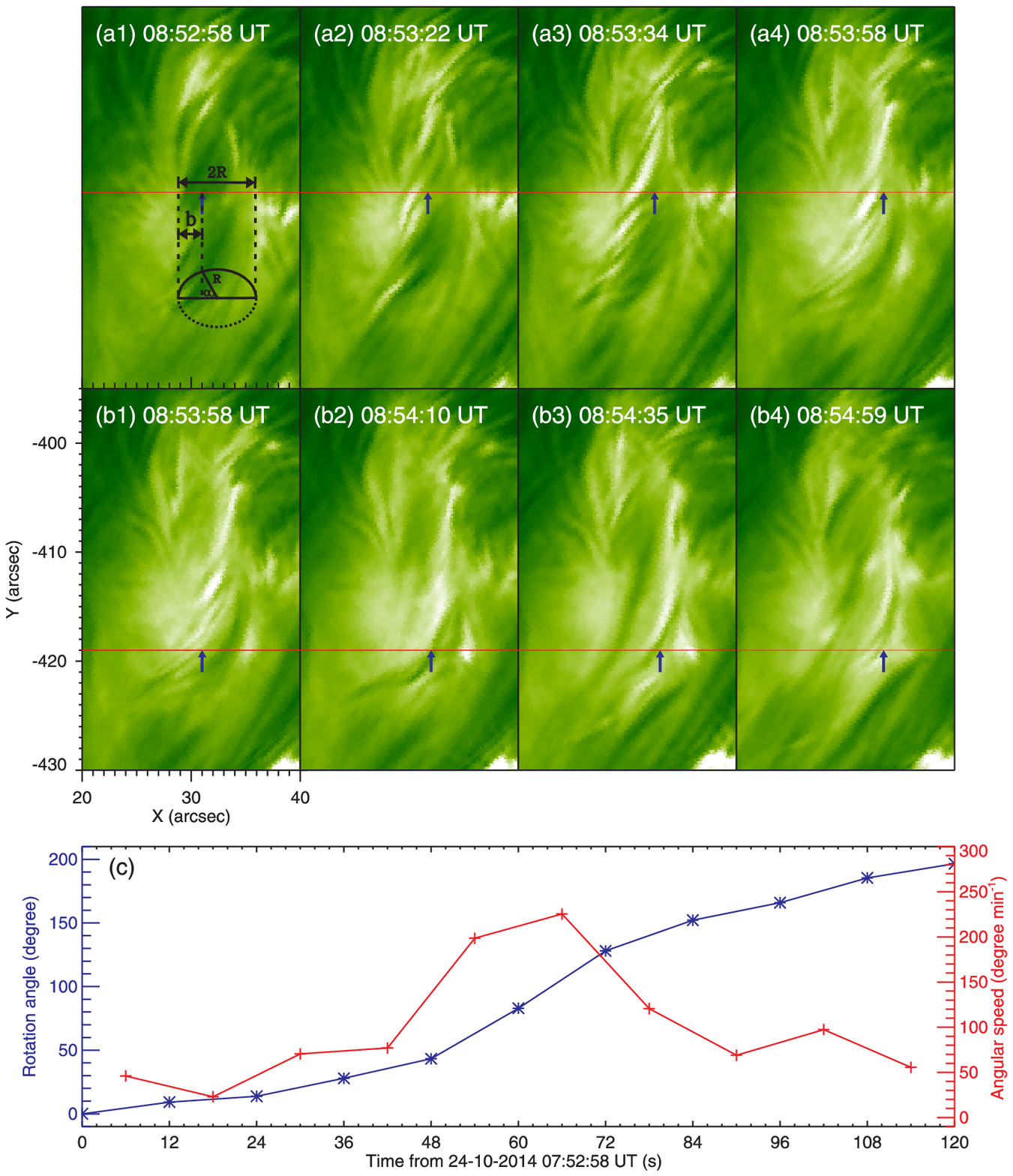}
\caption{Panels (a1)-(a4) and (b1)-(b4): NVST H$\alpha$ images
displaying the rotation of the filament. The red straight lines are
the reference lines. The blue arrows indicate the intersection
positions of the filament threads and the red lines. Panel (c):
variations of the rotation angle and angular speed of the filament
in two minutes. The inserted cartoon in panel (a) is used
to illustrate how we measure the rotation of the filament. \label{fig5}}
\end{figure*}

The NVST at the \emph{Fuxian Solar Observatory} has three channels
to image the Sun, i.e., H$\alpha$, TiO, and the G band. The
H$\alpha$ channel is used to observe the solar chromosphere with a
bandwidth of 0.25 {\AA}. The H$\alpha$ data used in this study were
obtained in 6562.8 {\AA} from 07:50:12 UT to 09:17:33 UT on 2014
October 24 with a cadence of 12 s. The NVST was pointed to AR 12192
with a field-of-view (FOV) of 152$\arcsec$$\times$157$\arcsec$ and a
pixel size of 0.$\arcsec$168. The data are first calibrated,
including flat field correction and dark current subtraction, and
then reconstructed using speckle masking (Weigelt 1977; Lohmann et
al. 1983).

We also adopt the Atmospheric Imaging Assembly (AIA; Lemen et al.
2012) multi-wavelength images and the Helioseismic and Magnetic
Imager (HMI; Scherrer et al. 2012; Schou et al. 2012) data on board
the \emph{SDO}. We choose the AIA 131 {\AA}, 171 {\AA}, 304 {\AA}
and 1600 {\AA} images, obtained from 07:00 UT to 09:20 UT on October
24 with a pixel size of 0$\arcsec$.6 and a cadence of 12 s. We use
the HMI full-disk line-of-sight (LOS) magnetograms with a 3 min
cadence, i.e., one frame in four, from 2014 October 23 12:00 UT to
2014 October 24 12:00 UT.

\section{Results}

By investigating the evolution of the HMI magnetograms, we found
that new magnetic flux started to emerge from 2014 October 21 in the
southeast of the main negative polarity sunspot in AR 12192. About 3
days later, a blowout jet occurred at the emerging flux region,
triggering an M4.0 flare and a CME. The flare and the associated jet
observed by the \emph{SDO} and NVST near the peak time were shown in
Figure 1. The GOES soft X-ray 1$-$8 {\AA} flux showed that the M4.0
flare initiated at 07:37 UT, and reached its peak at 07:48 UT (see
the red curve in panel (b)). As shown in panel (a), the east flare
ribbon was located at the negative polarity fields and the west
ribbon was at the positive polarity fields. The appearances of the
jet and the filament eruption were similar in higher-temperature
wavelength of 131 {\AA} and lower-temperature wavelengths of 171
{\AA} and 304 {\AA} (panels (b)-(c) and (e)). The two ribbons of the
flare in the chromosphere were observed in 1600 {\AA} image (panel
(d)). The jet (the bright structure denoted by arrows ``A") and the
filament (the dark structure indicated by arrow ``B") were clearly
observed in the H$\alpha$ image (panel (f)). Unfortunately, the
observations of NVST started late and we can only use the \emph{SDO}
data to study the first eruption.

The temporal evolution of the first eruption is shown in Figure 2
(also see the accompanying animation movie\_1.mp4). The filament was
in the east-west direction, with a projected length of about 60 Mm.
The east footpoint of the filament was anchored in the negative
polarity magnetic fields and the west one is at the positive
polarity fields (panel (a)). The blowout jet appeared at about 07:38
UT, and the material of the bright features was simultaneously
ejected and moved towards the south (indicated by the arrows in
panels (b)-(c)). The most distinct change of the filament is that it
began to rise (denoted by the arrow in panel (d)). From 07:44:19 UT,
the projected height of the filament increased by about 5 Mm in two
minutes. The region with EUV brightenings was extended as seen in
the AIA 304 {\AA} images and the flare reached its peak at 07:48 UT
(panel (e)). Some filament material was ejected outward (denoted by
the arrow in panel (f)). At 07:59 UT, the flare had greatly decayed
(panel (g)), and the filament erupted (indicated by the arrow in
panel (h)). At the decaying phase of the flare, partial material of
the erupting filament fell back to the solar surface nearby its east
footpoint. At about 08:12 UT, a CME was observed in the FOV of LASCO
C2. The angular width of the CME was about 90\degr and the average
velocity was approximately 500 km s$^{-1}$.

Figure 3 displays the second eruption process of another small
filament observed by AIA 171 {\AA} and 304 {\AA}. The general
appearance and the surrounding environment before the second
eruption is shown in panel (a). The second filament erupted at
(40$\arcsec$, -430$\arcsec$) about twenty minutes after the first
filament eruption. We can see that another blowout jet was
associated with the filament eruption. The blue arrows in panels
(b1) and (c1) point to the filament before eruption. A set of
``open" field lines were rooted in vicinity of the east footpoints
of the filament, both anchoring in the negative polarity magnetic
fields. The ``open" lines (denoted by the arrows in panel (a)) were
outlined by the falling dark material of the former jet. The west
footpoint of the filament was shielded by the material of the dark
features of the ``open" lines, and seemed to be located at the
positive polarity fields. At about 08:41 UT, the interaction between
the ``open" lines and the filament began, and meanwhile the EUV
brightenings were enhanced (panels (b2) and (c2); also see the
animation movie\_2.mp4). The interaction continued and then the
filament erupted finally (panels (b3) and (c3)).

With the high-resolution observations of the NVST, the detailed
evolution process of the second eruption was analyzed (see the
accompanying animation movie\_3.mp4). The filament and the ``open"
lines are denoted in Figure 4(a) by arrows ``A" and ``B",
respectively. By comparing the H$\alpha$ and 171 {\AA} images, we
find that the structure denoted by arrow ``B" in panel
Figure 4(a) corresponds to the ``open" lines indicated by
the blue arrows in Figure 3(a). Therefore the structure ``B" in
Figure 4(a) represents a set of ``open" lines. The footpoint of the
``open" lines was located at the negative polarity magnetic fields,
lying at the northwest of the filament (Figure 4(a)). Then the
``open" lines began to interact with the filament and they gradually
reconnected. At 08:40:03 UT, many ``open" lines within the area
marked by the brackets disconnected from their initial footpoints
(panel (b)). The ``open" lines continued moving eastward and
reconnecting with the filament. During the reconnection process, the
interaction area of the filament and the ``open" lines was
brightened up as observed in EUV lines, e.g., AIA 171 {\AA} line
(denoted by the green arrow in Figure 3(b2)). At 08:43:53 UT, the
``open" lines totally connected with the filament, and a new system
of the filament and ``open" lines was formed, as displayed in panel
(c). After that, the filament started to untwist, triggering the
rotation of the new system and leading to the brightening (panel
(d)). The highly twisted system rotated continuously (panel (e)),
and lasted for about twenty minutes. The rotation angle caused by
the untwisting motion was about 8$\pi$ (see next paragraph for
detail). At the late phase of the rotation, the twist of the system
was significantly released (panel (f)). The comparison of panels (a)
and (f) showed that the footpoints of the ``open" lines were shifted
to the east end of the filament. As the system rotated, the magnetic
twist propagated outward along ``open" lines (see the accompanying
animation movie\_4.mp4). To show the propagation process well, we
take a bright helical structure as an example. At 08:53:35 UT, the
projected position of the helical structure was located at
(40$\arcsec$, -475$\arcsec$), as denoted by the arrow in panel (g).
About four minutes later, the helical structure reached to a new
position (25$\arcsec$, -600$\arcsec$) (indicated by the arrow in
panel (h)). The average moving velocity of the structure was
estimated to be about 400 km s$^{-1}$.

For detailed analysis of the untwisting process of the filament and
the ``open" lines, we focus on the marked rectangle regions in
Figure 4(e), track the shift movement of dark fine structures, and
mark the position changes of the structures with time (see Figure
5). The dark threads in the H$\alpha$ images are considered to
represent flux tubes of the filament. In order to measure the
rotation angle of the filament, we draw a red line which is
perpendicular to the filament spine. As the filament rotated, the
intersection (marked by the blue arrows) between each dark thread
and the red line shifted from left to right (panels (a1)-(a4) and
(b1)-(b4)), indicating a clockwise rotation. We deduce the rotation
angle from the shift distance. To illustrate how we measure
the rotation of the filament, a cartoon is inserted into panel (a).
The filament is considered to be a circular cylinder. The width of
the filament is $2R$ and the distance from the intersection to the
filament left edge is $b$. We define $\cos \alpha = (R-b)/R$, and
then the rotation angle $\Delta \alpha$ between $t_{1}$ and $t_{2}$
is $\alpha _{2} - \alpha _{1}$. This measurement method works under
the assumption that the filament consisting of dark threads rotates
rigidly. Because we do not trace and measure the movement of dark
features along the field lines, the material flow will not affect
our results. For one analyzed thread, it was on the left edge of the
filament at the beginning (panel (a1)). About 24 seconds later, the
intersection position moved to the west for about 3$\arcsec$ (shown
in panel (a2)). Until 08:53:34 UT, the intersection moved about
6$\arcsec$ (panel (c)). We calculate the movement of the
intersecting position every 12 seconds, until the thread moved to
the right edge of filament (indicated in panel (a4)). Since this
thread could not be observed at the far side of the filament, we
choose another thread and track its evolution (marked by the arrows
in panels (b1)-(b4)). Thus we can obtain the rotation angle and
angular speed of the filament in two minutes, as plotted in panel
(c). The angular velocity is not uniform, revealing the acceleration
process followed by a deceleration. The minimum and maximum angular
speeds are 30{\degr} min$^{-1}$ and 230{\degr} min$^{-1}$,
respectively. The average angular velocity is 100{\degr} min$^{-1}$.
The rotation process lasted for about twenty minutes, the total
twist angle of 8$\pi$ is reasonable since the angular velocity is
not uniform. Generally, many mechanisms can result in the
eruption of flux ropes, such as kink instability (T{\"o}r{\"o}k \&
Kliem 2005), and torus instability (Kliem \& T{\"o}r{\"o}k 2006).
However, according to some previous studies, even for large twists,
low-lying filaments, especially those with small length, still can
keep stable (e.g., Vrsnak 1990; Isenberg \& Forbes 2007).

\section{Conclusions and Discussion}

With the \emph{SDO} observations, we study the only one
CME-associated flare in AR 12192. There were two filament eruption
processes at the edge of the AR. The magnetic flux emerged
continuously at the southeast of the negative main sunspot in this
super AR, forming a twisted filament system at this area. The first
filament eruption caused a blowout jet, and produced an M4.0 flare
and led to the only one CME. Using the NVST high tempo-spatial
resolution data, we particularly study the evolution of the second
filament eruption. The filament and the ``open" lines reconnected
with each other and formed a new system, the details of which were
clearly observed by the NVST. Subsequently, the filament untwisted,
and resulted the rotation of the new system. The total rotation
angle is about 8$\pi$ and the average angular speed is about
100{\degr} min$^{-1}$.

As the largest AR since 1990 November, NOAA 12192 is the most flare
productive AR. Except for one M4.0 class flare, all the X-flares and
the other M-flares which originated from the AR core were not
associated by any CME. According to the recent studies about this
AR, the confined flares were affected by the strong confinement of
the overlying magnetic fields, and the decay index of the background
fields was below the typical onset threshold for the torus
instability and thus no CME was produced (Sun et al. 2015; Thalmann
et al. 2015; Chen et al. 2015). However, the M class flare in the
present study was located at the edge of the AR instead of in the
core region, and a CME was produced. We note that this flare was
triggered by the blowout jet, and the flare became an eruptive flare
due to the interaction between the flare material and
``open" fields, leading to the CME (Thalmann et al. 2015).

In the standard jet model, the jet is caused by the reconnection
between the emerging small-scale loops and the pre-existing ``open"
lines. While according to the blowout jet model, there exists a
filament which is involved into the reconnection and eruption.
Although many authors (e.g., Hong et al. 2011; Shen et al. 2012;
Pucci et al. 2013) have reported the observational evidences for the
later model, the interaction process between the filament and the
``open" field lines have not been well observed. In this paper, the
filament can be identified in multi-wavelengths, such as AIA 304
{\AA}, 171 {\AA}, and 131 {\AA}, and especially in the NVST
H$\alpha$ line. As revealed by the H$\alpha$ image in Figure 4(a),
the location of one end of the filament was close to the ``open"
lines. Then the interaction between them began (Figure 4 (b)). About
16 minutes later, their topology had been significantly changed, and
the new connection between the filament and the ``open" lines was
created (Figure 4 (c)). Thanks to the high-quality H$\alpha$ data
from the NVST, this detailed interaction process is, to our
knowledge, first reported in the present paper.

As pointed out by Zhang \& Low (2005), the accumulation of magnetic
helicity in the corona can build up magnetic energy which will be
responsible for the CME. When the stored free energy is higher than
the Aly limit (Aly 1984, 1991), eruptions take place. As a
fraction of helicity in a system, magnetic twist can be inferred by
filament structure. Thus we try to study magnetic twist through
observing the evolution of filament structures. In the present
study, we show a scenario of twist transfer process between
different systems and propagation from the lower atmosphere to the
higher layers. As shown in Figure 4(a), the twist was initially
stored in the helical filament. After the interaction between the
filament and the ``open" lines, a new system connecting the filament
and the ``open" lines was formed. Then the new system rotated due to
the untwist of the filament, and thus the twist was transferred from
the closed filament system to the ``open" system. During the
rotating movement, the twist probably propagated from the lower
atmosphere to the upper layers. We speculate that the twist might be
finally spread by the CME to the interplanetary space.

\acknowledgments { We thank the referee for valuable suggestions.
This work is supported by the National Natural Science Foundations
of China (11203037, 11533008, 41331068, 11303050, and 11221063), the
CAS Project KJCX2-EW-T07, the National Basic Research Program of
China under grant 2011CB811403, the Strategic Priority Research
Program$-$The Emergence of Cosmological Structures of the Chinese
Academy of Sciences (No. XDB09000000), and the Youth Innovation
Promotion Association of CAS (2014043). The data are used courtesy
of NVST, HMI, and AIA science teams. \\}

{}

\end{document}